# Anomalous temperature-dependent spin-valley polarization in monolayer WS$_2$

A.T. Hanbicki[1], G. Kioseoglou[2,3], M. Currie[1], C.S. Hellberg[1], K.M. McCreary[1], A.L. Friedman[1], and B.T. Jonker[1]

[1]*Naval Research Laboratory*, Washington, DC 20375
[2]*University of Crete*, Heraklion Crete, 71003, Greece
[3]*Institute of Electronic Structure and Laser (IESL), Foundation for Research and Technology Hellas (FORTH),* Heraklion Crete, 71110, Greece

Single layers of transition metal dichalcogenides (TMDs) are direct gap semiconductors with nondegenerate valley indices. An intriguing possibility for these materials is the use of their valley index as an alternate state variable. Several limitations to such a utility include strong, phonon-enabled intervalley scattering, as well as multiparticle interactions leading to multiple emission channels. We prepare single-layer WS$_2$ such that the photoluminescence is from either the neutral or charged exciton (trion). After excitation with circularly polarized light, the neutral exciton emission has zero polarization, however, the trion emission has a large polarization (28%) at room temperature. The trion emission also has a unique, non-monotonic temperature dependence that we show is a consequence of the multiparticle nature of the trion. This temperature dependence enables us to determine that coulomb assisted intervalley scattering, electron-hole radiative recombination, and a 3-particle Auger process are the dominant mechanisms at work in this system. Because this dependence involves trion systems, one can use gate voltages to modulate the polarization (or intensity) emitted from TMD structures.

In low-dimensional, hexagonal lattice structures there are two degenerate yet inequivalent K-points in the Brillouin zone, labeled K and K'. These high symmetry points are at the edge of the Brilloun zone and are usually local extrema: a local maximum in the valence band and a local minimum (or valley) in the conduction band. Much like electron spin, this valley index can be used as a new state variable to control the operation of an electronic device [1,2,3]. Single layers of transition metal dichalcogenides (TMDs) are semiconductors with a direct gap at the K-points, and are well known for their potential as valleytronic materials [4,5,6]. One of the ways to access the valleytronic functionality in these materials is via spin-valley coupling. Because of strong orbital hybridization and time-reversal symmetry, the valence band maximum in each valley has only one spin state (the conduction band is nearly spin degenerate), giving these materials unique optical selection rules [4,5,6,7]. It is therefore possible to selectively populate and interrogate the different valleys, K or K', using circularly polarized light. There has been much experimental work exploring the spin-valley coupling with optical techniques [4,5,6,8,9,10,11,12], and some recent progress coupling these optical techniques with transport measurements to hint at possible applications for these materials [13,14].

Optical excitation creates electron-hole pairs that can bind to form quasiparticles known as excitons. When TMDs are excited with circularly polarized light, excitons are created in a single valley. The radiative decay of excitons within this valley subsequently produces circularly polarized light due to optical selection rules. Therefore, measuring the circular polarization of electroluminescence or photoluminescence (PL) provides a direct way to monitor valley populations. These populations are altered by intervalley scattering, a process that is enabled by phonons which conserves momentum [10,12,15]. At high temperatures or high photoexcitation energy, large phonon populations can readily couple the valleys, reducing valley specific populations. Because of this strong intervalley coupling, exciton polarization is often only seen in systems measured at low temperature or near resonant pumping conditions [6,8,9,10].

In addition to the excitonic emission, photoluminescence spectra from TMDs have a plethora of other features that speak to the complexity and richness of these systems [6,11,16,17]. For instance, a third carrier (electron or hole) can bind to the exciton complex creating a charged exciton, or trion, with an emission energy lower than the exciton because of the binding energy of the trion. Unfortunately, this adds a confounding dimension to the interpretation of the results, and indeed the utility of TMDs as valleytronic materials because the origin of many of the emission features is unknown and they are difficult to create reproducibly. For instance, in $WSe_2$ varying degrees of polarization have been measured for various emission channels of unknown origin [11].

Here we show that in naturally n-doped, single layer $WS_2$ we can isolate two distinct initial states: a neutral exciton and a trion. When excited with circularly polarized light, the neutral exciton has zero circular polarization in the PL at room temperature, however the trion has a polarization as high as 28%. Even with excitation energies far from the resonant condition (>300 meV), the trion emission maintains some circular polarization. The trion emission polarization also displays a unique temperature dependence. There is a pronounced increase in polarization at an intermediate temperature with a broad peak of 42% from 175–250 K. We show that non-monotonic temperature dependence is a consequence of the multiparticle nature of the trion and propose a second recombination channel that is more readily available in 3-particle systems. This enables us to determine the dominant mechanisms at work in this sub-system and to speculate on the suitability of this material for future valleytronic applications.

**Results**

**Initial Conditions.** Samples used here include monolayer flakes mechanically exfoliated from bulk $WS_2$ single crystals as well as large area layers of $WS_2$ grown by chemical vapor deposition. A microscope image of a representative exfoliated sample is shown in Figure 1a, and the Raman spectrum of the monolayer is shown in Figure 1b. The energy separation of the two Raman modes and accompanying PL confirm the single layer nature of our sample. Additional details of sample preparation are found in the methods and supplementary information. The data presented below is from this representative sample, but all of the results were reproduced on several monolayer samples of independent origin.

Once the single layer regions are identified, we measure the normalized reflectivity (Figure 1c), and the energy-resolved PL (Figure 1d) at 300 K. For $WS_2$, these spectra depend crucially on the preparation of the sample. In Figure 1c and 1d, the as-exfoliated spectra (thin, blue lines) are distinct from those measured after purposeful preparation in vacuum (thick, red lines). The sample preparation [18] consists of rastering a 532 nm laser (2 mW power and ~1μm spot size) across the entire flake while in a $10^{-6}$ Torr vacuum to desorb weakly bound contaminants [19,20]. As can be seen in Figure 1c and 1d, after treatment a peak emerges 33 meV below the peak observed in the as-exfoliated sample. The n-type conductivity measured by transport increases significantly after treatment as well. More details of this effect are described in the methods, supplemental materials, and elsewhere [18].

When excess electrons are present, the neutral exciton ($X^0$) can capture an electron to form a negatively charged exciton, or trion ($X^-$). In GaAs, for instance [21], at a critical electron density there is a sharp transfer of oscillator strength from $X^0$ to $X^-$, and eventually $X^0$ is completely quenched. A similar behavior is seen here. We attribute the high energy feature in Figure 1c and 1d to $X^0$ (thin, blue line), and the low energy feature to $X^-$ (thick, red line). An identical assignment was made in reference [20].

The separation of the $X^0$ and $X^-$ peaks in the reflectivity spectra is 33 meV. This energy is consistent with the binding energy of the charged exciton measured for $MoS_2$ [16] and MoSe2 [17], and predicted for all of the monolayer transition metal dichalcogenides [22]. In our PL measurement, the $X^-$ energy is shifted to a lower energy than observed in reflectivity. Such a shift is commonly seen for trions and can be attributed to a bandgap renormalization due to the electron density [21]. The high electron density also causes the width of the trion to be much larger than the free exciton. Indeed, the width of the trion PL can be used to estimate the doping level [21]. Based on our low temperature spectra, we estimate the electron density to be $5 \times 10^{12}$ $cm^{-2}$. In a material like GaAs, such a dense electron gas would lead to a breakdown of the trion and result in a Fermi-edge singularity. In the TMD systems, the binding energy of the exciton [23,24,25,26,27] and the trion [17,22] are much larger than traditional materials like GaAs [28], so we expect and confirm that the trion is stabilized for a much larger electron density.

**Measuring the Polarization.** From Figures 1c and 1d it is clear that we are able to isolate the $X^0$ and $X^-$ peaks by conditioning the sample. This allows us to reproducibly prepare the surface so that trends in temperature, excitation energy, and circular polarization can be reliably measured for each [18]. Figure 2 shows the PL of $WS_2$ from the neutral exciton (Figure 2a), and from the trion, (Figure 2b) at room temperature (left panels) and 4 K (right panels). The spectra were obtained with a circularly polarized excitation source with positive helicity (σ+) and excitation energy of 2.087 eV (594 nm laser). The resulting emission was analyzed for positive (σ+, solid red line) and negative helicity (σ–, open, blue circles). The polarization is defined as $P = [\ I(\sigma+)$

− $I(\sigma-)$ ] / [ $I(\sigma+)$ + $I(\sigma-)$ ], where $I(\sigma\pm)$ is the emission intensity analyzed for positive (negative) helicity. The most notable feature of these spectra is that, even at room temperature, the trion has a very large circular polarization, $P = 28\%$. This is in marked contrast to the free exciton, which has 0% polarization at room temperature using the same excitation conditions. The results at low temperature are also unexpected. The neutral exciton has a polarization roughly half that of the trion, and the trion polarization is about the same at low temperature as it is at room temperature. [29]

To understand the origin of the large room temperature polarization of the trion, as well as the relatively meager polarization at low temperature we measured the helicity-resolved PL from prepared $WS_2$ as a function of temperature for 10 different excitation energies. Figure 3a and 3b are representative sets of data taken with positive helicity excitation sources of 2.087 eV (594 nm), and 2.331 eV (532 nm), respectively. For each excitation energy we obtained data from 4 K to 300 K, and in these figures the spectra are offset for clarity. Plots of spectra from all ten excitation energies are provided in the supplemental information.

**Analyzing the Polarization.** A compilation of the raw data for the trion polarization is displayed in Figure 3c. In this figure the polarization is plotted as a function of temperature for all the excitation energies used. There are several interesting and novel features in these data.

First, the trion polarization at room temperature is 28% for excitation 150 meV above the emission energy. As the temperature increases or the excitation energy increases, phonons become available and the polarization decreases rapidly because of intervalley scattering [10]. Several contradictory room temperature polarization measurements have been reported for $MoS_2$ [6,9]. Mak and coauthors report zero polarization at room temperature, while Sallen, et al. report 40% polarization at room temperature. In both cases, they used an excitation energy 100 meV higher than the emission energy. To our knowledge no room temperature polarization has been reported for any other TMD.

Second, at low temperature (4 K), the polarization never exceeds 25%, even with an excitation energy only 100 meV above the emission energy. When the excitation energy is less than 2 LA phonons above the emission energy, there is insufficient energy to excite the phonons necessary for intervalley scattering, and the polarization should be 100% [10]. For $WS_2$, 2 LA is 46 meV [30,31], for $MoS_2$, it is 60 meV [30]. Although we are exciting the $WS_2$ near the emission energy, we observe a relatively modest polarization of 25% at 4 K. This is in marked contrast with the very high polarizations reported at low temperature for $MoS_2$ [6,9].

The most novel feature of these data is that the polarization *increases* significantly with increasing temperature, even when exciting the system with an energy 300 meV above the emission energy. Such an increase in the polarization at intermediate temperature has not been observed before, and we propose that this is a unique consequence of the emission originating from a 3-particle entity (the trion) rather than from a simple exciton. When the excitation energy is 350 meV higher than the emission energy (i.e. laser wavelengths < 532 nm), the polarization decreases smoothly as a function of temperature as observed in other TMD monolayers [10].

To elucidate the mechanisms leading to the anomalous temperature dependence, we focus on the data collected using the lowest excitation energies, i.e. near-resonant excitation (within 150 meV of the exciton emission). Figure 4a shows the temperature dependence of the circular polarization using the 2.087 eV and 2.109 eV sources (594 nm and 588 nm). The solid lines, guides to the eye, show two clear polarization levels. At low temperature (solid blue line) the polarization is 25%. At 120 K (~10 meV) the polarization begins to increase steadily until it plateaus (solid red line) at 42% above 175 K (~15 meV). Above 275 K (~24 meV) the polarization begins to decrease. This eventual decrease is due to a combination of effects including enhanced intervalley scattering from phonons, the dissociation of the trion, and the spin-orbit split valence band, and are outside the realm of our discussion.

**Discussion**

While it is fairly straightforward to imagine scenarios where the polarization decreases with increasing temperature, it is difficult to explain an *increase* in polarization with increasing temperature. We use a rate equation framework, developed earlier [6,10] to understand the low polarization observed at low temperature as well as the origin of the increase in polarization. In this approach, we consider the time evolution of the carrier populations in the K and K' valleys. By exciting the system with circularly polarized light we first create free excitons in a single valley (K, for instance). Next, trions rapidly form due to the high electron density. In a simple, single particle picture, the trions can form either with both electrons in a single valley, or one electron in each valley. Electron-hole recombination and intervalley scattering then governs the evolution of the system.

In the steady state, the observed polarization in 2-dimensions is [6,10]

$$P = \frac{P_0}{\left(1 + 2 \cdot \beta/\alpha\right)} \quad . \tag{1}$$

Here $P_0$ is the initial polarization of the system, $\alpha$ is the exciton radiative recombination rate, and $\beta$ is the rate of spin relaxation. Note that $\alpha = 1/\tau_r$ and $\beta = 1/\tau_s$ where $\tau_r$ is the recombination time, and $\tau_s$ is the spin relaxation time. Because of the unique selection rules in this system [4,5,6,7], spin relaxation is related to the intervalley scattering rate. To first order, we consider the spin relaxation time equivalent to the intervalley scattering time. A schematic definition of these processes as well a full derivation of this equation is presented in the supplemental information. Some insight into the physical processes of this system is obtained when we consider the relation of these characteristic rates and times.

At low temperature, the exciton recombination rate, $\alpha$, is simply the light-emitting electron-hole recombination event in a single valley. The intervalley relaxation, $\beta$, is the Coulomb-mediated exciton spin flip-flop where an exciton scatters from one valley to the other. [6,12,32] The intervalley scattering process is fast if the exciton has a large center of mass momentum, and slower if the exciton is close the ground state momentum. [11,33] For excitons generated by photons far from resonance, i.e. with high energy, we expect a large initial center of mass momentum which will be reduced as they relax. Capturing an electron to form a trion will also quickly reduce the momentum. Since we are exciting the system with circularly polarized light, the initial polarization of the system is expected to be very high, i.e. $P_0 \rightarrow$ 100%. A polarization, $P$, considerably less than this initial polarization means the intervalley relaxation, $\beta$, must be faster than the exciton recombination rate, $\alpha$, and/or the initial polarization, $P_0$, could somehow be reduced, according to equation (1). Our observed trion polarization of 25% at low temperature means the ratio $\tau_r/\tau_s = \beta/\alpha$ is 1.5 assuming that the initial polarization is 100%. This initial polarization is likely reduced, however, since we are measuring the trion polarization. As is seen in Figure 2, the polarization of the neutral exciton is roughly half that of the trion. Since the trion forms via excitons capturing electrons, any relaxation prior to trion formation will reduce $P_0$ for the trion. Indeed, if $P_0$ is 50%, the ratio $\beta/\alpha$ is 0.5, still the same order of magnitude. Recombination lifetimes on the order of 1~5 psec have been measured in other TMDs [34], and suggests the intervalley scattering lifetime must be of the same picosecond timescale.

Another important recombination channel that should be considered is the non-radiative Auger process. Auger recombination has been found to be strong in $MoS_2$ monolayers. [35,36] In this three-particle process, an electron and hole recombine non-radiatively, transferring energy to a third electron (or hole), which moves to a higher energy level. [37] The Auger recombination rate of trions has been observed to increase with increasing temperature in both nanocrystals [38] and in $MoS_2$ monolayers. [39] The non-radiative Auger rate, $A$, can be incorporated into the rate equation analysis and yields a polarization

$$P = \frac{P_0}{\left(1+2\cdot\beta/(\alpha+A)\right)}. \tag{2}$$

The addition of a second recombination channel has consequences on the light emission intensity as well. The resulting intensity is given by,

$$I_{tot} = I(\sigma+) + I(\sigma-) = \frac{\alpha}{(\alpha+A)} \tag{3}$$

where again $I(\sigma\pm)$ is the measured intensity of positive (negative) helicity light. It is clear that an increase in the Auger recombination rate, $A$, will simultaneously cause a drop in the measured intensity along with an increase in the measured polarization. Indeed, as is shown in figure 4b, there is a marked drop in emission intensity from 125 K to 175 K, the same temperature range where there is an increase in polarization. More quantitatively, the temperature dependence of the intensity can be fit with a single exponential function, $I = I_o e^{-cT}$ where $c$ is a constant. Using the same exponential decay constant, $c$, and only changing the intensity prefactor, $I_o$, the low temperature fit (solid blue line) and high temperature fit (solid red line) are related by $I_{o\text{-}low}/I_{o\text{-}high}$ = 4. This ratio suggests the Auger rate is 3 times larger than the radiative recombination rate based on equation (3). Therefore, in this model, if $A = 3\alpha$, $\beta = 0.5\alpha$, and $P_0 = 50\%$, then at low temperature the polarization should be 25% and at high temperature it should jump to 40%.

These values sufficiently reproduces the trends observed in our data to suggest the Auger process is a significant effect in the high temperature behavior of $WS_2$ trion system.

In summary, we have prepared naturally n-doped, single-layer $WS_2$ such that the emitted PL is from either the neutral exciton or the trion. The measured degree of circular polarization shows that while the neutral exciton has zero polarization at room temperature, the trion exhibits a polarization of 28%. The trion polarization also exhibits a distinct, non-monotonic behavior with temperature – the polarization has a broad peak of 42% between 175 and 250 K. To explain this anomalous behavior, we develop a model that includes a 3-particle recombination mechanism. Coulomb assisted intervalley scattering, electron-hole radiative recombination, and a 3-particle Auger process are the dominant mechanisms at work in this system and account for the novel temperature dependence. Because this dependence is unique to the trion systems, one can use, for example, a gate voltage to switch the polarization (or intensity) emitted from these TMD structures. The circular polarization modulation could be used to control interactions between chiral materials on a sub-micrometer scale, enabling various valleytronic applications/systems.

## Methods
**Sample synthesis and isolation.** Three different samples were used in this study: an exfoliated monolayer purchased from 2D semiconductors; a monolayer that we exfoliated from a bulk crystal; a large-area, single monolayer grown by chemical vapor deposition (CVD). Further details of each is given in the supplementary information. Single layer regions are identified and confirmed in several ways. First, we sweep a 1 mW, 532 nm laser over the sample at room temperature. The PL from a $WS_2$ monolayer is orders-of-magnitude greater than from multilayers. In fact, luminescence from a monolayer is easily seen with a standard charge-coupled device camera as is shown in the inset of Figure 1a. Next, we measure the Raman spectra of the thin, optically active regions (Figure 1b). An energy separation of 60 cm$^{-1}$ between the in-plane $E^1_{2g}$ and out-of-plane $A_{1g}$ mode is a clear signature of a single layer of $WS_2$ [31].
**Sample preparation.** On all samples, exposing the sample to even a low power excitation source will start to modify the photoluminescence spectrum. The trion feature emerges when the sample is under vacuum and < 1 μW of laser power is applied to the sample. Therefore, preparation of the sample consisted of rastering a 532 nm laser of 2 mW power and ~1μm spot size across the entire flake while in a 10$^{-6}$ Torr vacuum. The exact result of this procedure is being investigated, however, we surmise that this treatment desorbs adsorbates from local heating [18]. A similar behavior was discovered on the other TMDs $MoS_2$, $MoSe_2$, and $WSe_2$ [16] however $WS_2$ seems especially susceptible to this effect. Indeed a similar phenomenon was observed in a preliminary study of $WS_2$ where different charge states were accessed by varying the excitation power [19]. Therefore, while local desorption due to laser heating is the most likely mechanism, photo-desorption of adsorbates or possibly photo-doping are also possibilities. The free exciton is completely and reproducibly recovered when the sample is exposed to air, or some oxygen containing species (not, for instance nitrogen or helium). Using this technique, we isolate the trion from the free exciton, and create a reproducible initial condition.

**Optical measurements.** We used a micro-Reflectivity/PL setup (spatial resolution of 1 μm) with a 50x objective, appropriate filters and incorporated a continuous-flow He-cryostat to collect reflectivity and PL in a backscattering geometry. Samples were excited with various continuous-wave lasers polarized as $\sigma^+$. Excitation energies are indicated where appropriate. Emitted light was dispersed by a single monochromator equipped with a multichannel charge coupled device (CCD) detector. The PL spectra were analyzed as $\sigma^+$ and $\sigma^-$ using a combination of quarter-wave plate (liquid crystal) and linear polarizer placed before the spectrometer entrance slit. We obtain the same polarization when the sample is excited with negative helicity light, and the emitted circular polarization is 0% when the sample is excited with linearly polarized light. The data at 4 K from the neutral exciton shown in Figure 2 was taken with a power of 0.7 μW and integration time of 2 seconds. Because the intensity decreases exponentially as a function of temperature, significant signal could not be collected for the neutral exciton at elevated temperature while the sample was in vacuum.

**Transport measurements.** Flakes were exfoliated onto $SiO_2$/n-Si substrates for the transport measurements. Top contacts were then deposited using standard e-beam lithographic techniques, and the n-Si was used as a global back gate. Channel resistance was monitored as a function of gate voltage to determine the sign of the charge carrier.

**Author Information**
**Corresponding author**
Hanbicki@nrl.navy.mil

**Acknowledgments**
This work was supported by core programs at NRL and the NRL Nanoscience Institute. We thank Jim Culbertson for assistance with Raman measurements. GK gratefully acknowledges the hospitality and support of the Naval Research Laboratory where the experiments were performed. KMM is a National Research Council Research Associate at NRL.

**Author Contributions**
A.H., G.K. and M.C. performed the experiments. A.H., G.K., M.C. and C.S.H. analyzed the data. A.F. performed transport measurements. A.F. and K.M.M. provide samples. All authors discussed the results and contributed to the manuscript.

**Additional Information**
Competing financial interests: The authors declare no competing financial interests.

**Figure captions**

**Figure 1. Monolayer WS$_2$ characterization**. (a) Optical microscope images of representative WS$_2$ flake with the monolayer region indicated. The inset shows the photoluminescence from the flake at room temperature to illustrate the spot size. (b) Raman spectrum of the monolayer regions taken at 300 K with an excitation energy of 488 nm. The splitting of the in-plane $E_{2g}^1$

mode and the out-of-plane $A_{1g}$ mode are characteristic of a single layer. Normalized (c) reflectivity and (d) photoluminescence spectra taken at room temperature for the as-deposited sample in air (thin, blue line) and the sample in vacuum prepared as described in the text (thick, red line).

**Figure 2. Polarization of monolayer WS$_2$ neutral and charged exciton.** Photoluminescence analyzed for positive (σ+: solid, red trace) and negative (σ–: blue, open circles) helicity of the (a) neutral exciton and (b) charged exciton. Spectra taken in the left (right) panels are taken at room temperature (4 K). The excitation was with an energy of 2.087 eV and positive helicity.

**Figure 3. Temperature and excitation energy dependence of the monolayer WS$_2$ charged exciton polarization**. Photoluminescence analyzed for positive (σ+: solid, red trace) and negative (σ–: blue, open circles) helicity as a function of temperature of the charged exciton with excitation energies of (a) 2.087 eV and (b) 2.331 eV. At each temperature, the spectra are normalized to the σ+ intensity and offset for clarity. (c) Summary of the circular polarization as a function of temperature for each excitation wavelength used.

**Figure 4. T**e**mperature dependent behavior of circular polarization and trion intensity.** (a) Temperature dependence of the circular polarization for the two highest excitation energies used (594 and 588 nm). The solid line is a guide to the eye to illustrate the two-level behavior. The Auger recombination rate, *A*, described in the text, is zero for the lower level of polarization and becomes non-zero after a certain temperature. (b) Intensity of the trion peak as a function of temperature. The solid lines through the data are fits assuming a simple exponential decrease in intensity with a low and high temperature prefactor.

Fig. 1

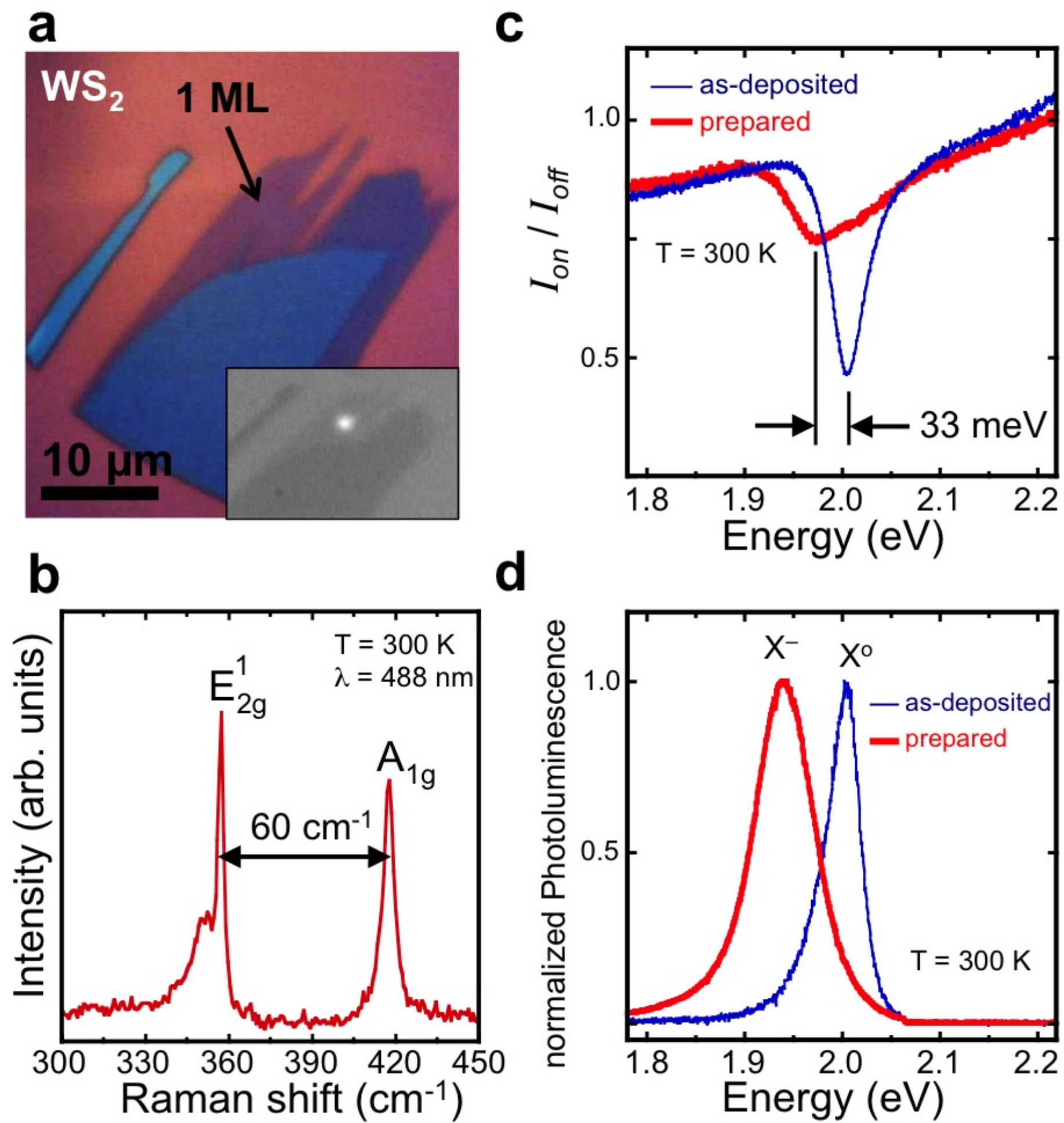

Fig. 2

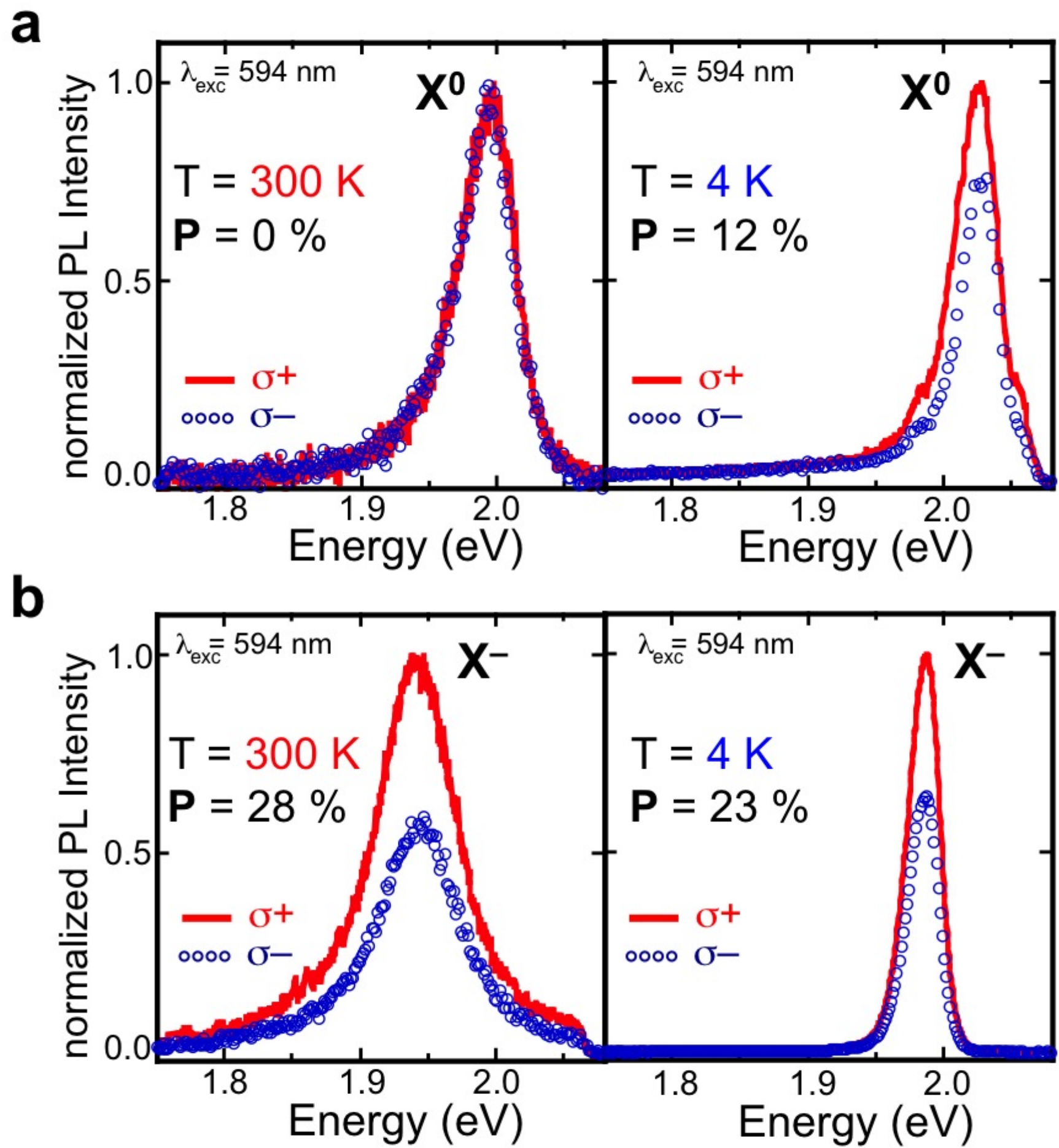

Fig. 3

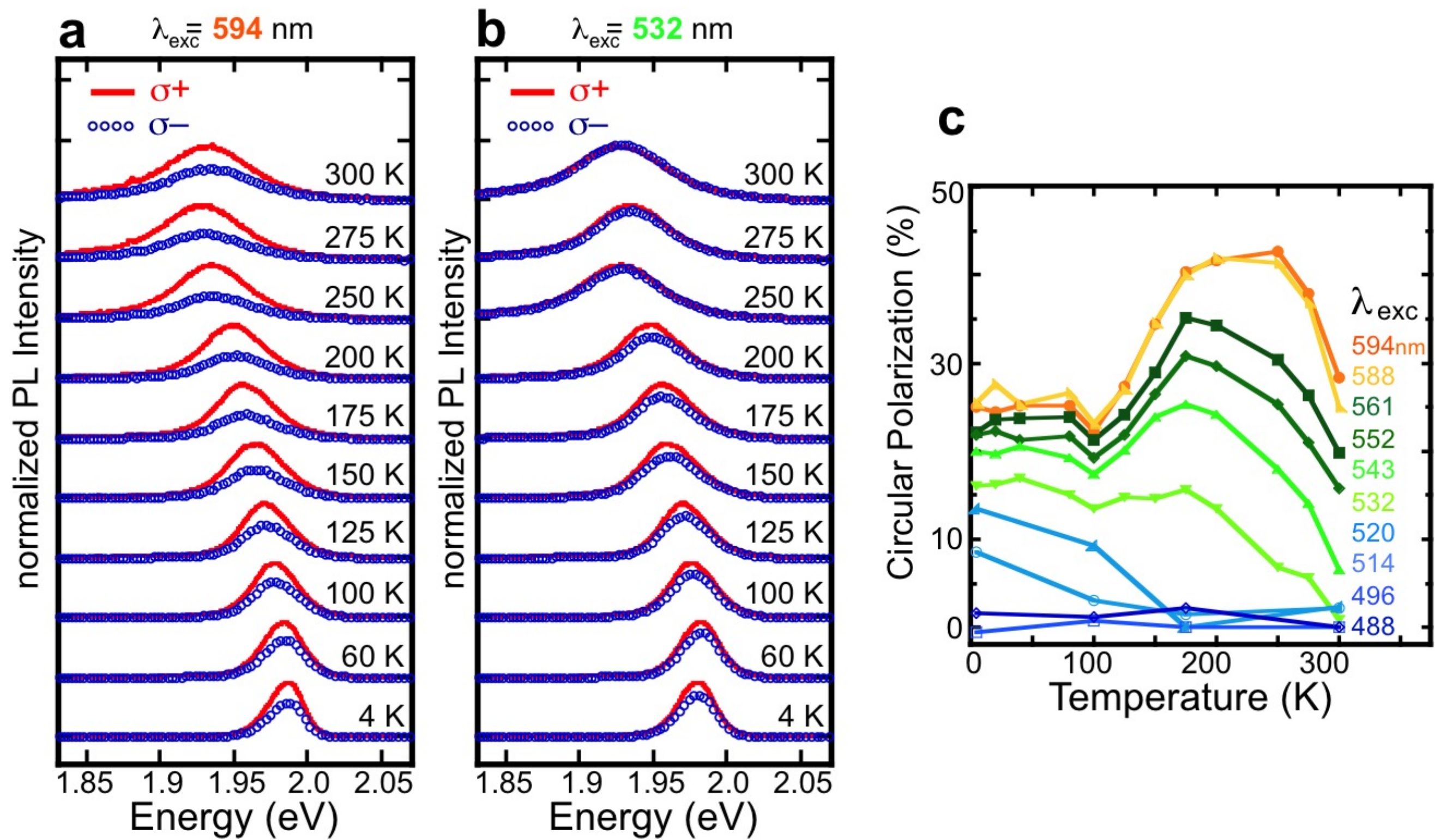

Fig. 4

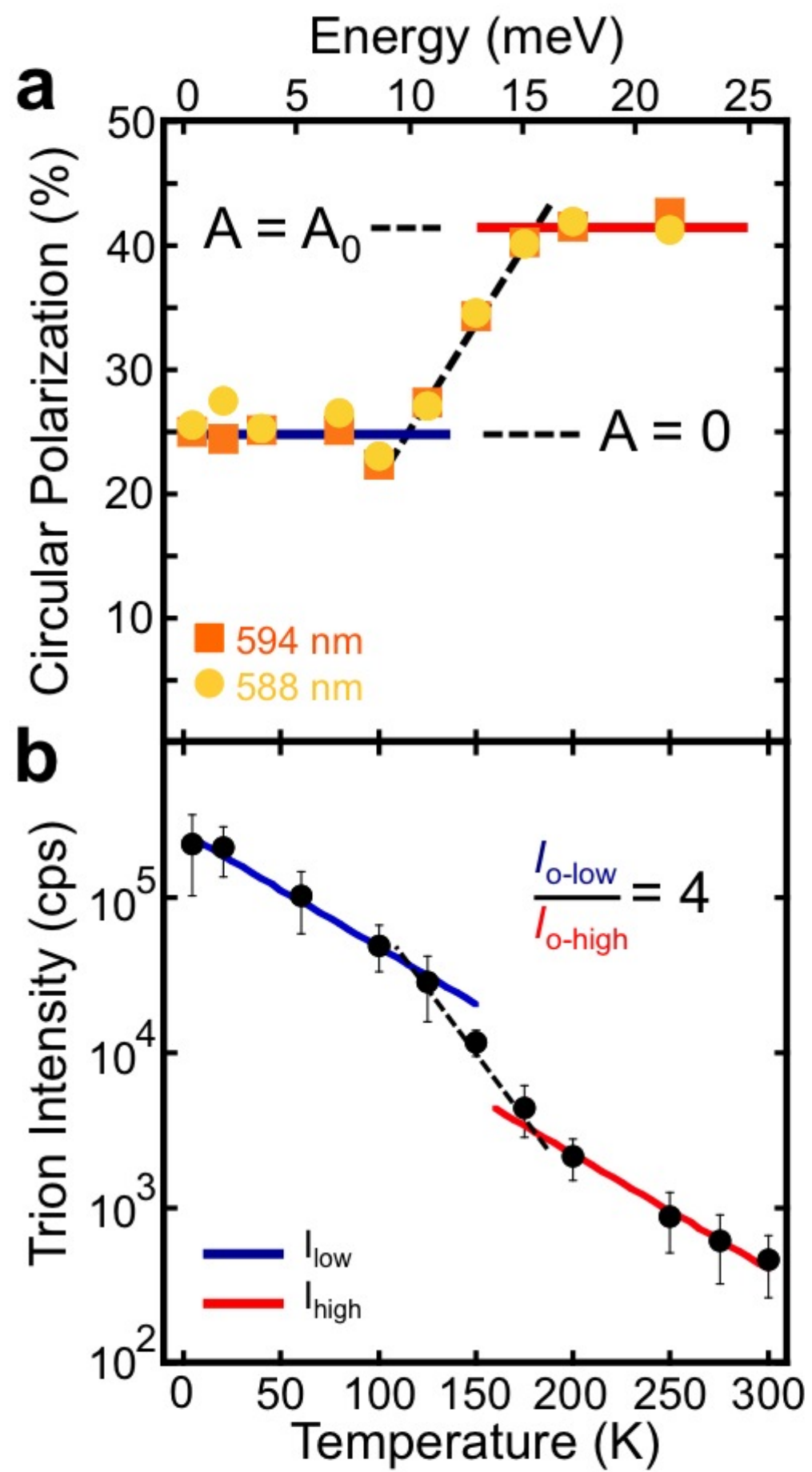

# Supplemental Information

# Anomalous temperature-dependent spin-valley polarization in monolayer $WS_2$

A.T. Hanbicki[1], G. Kioseoglou[2,3], M. Currie[1], C.S. Hellberg[1], K.M. McCreary[1], A.L. Friedman[1], and B.T. Jonker[1]

[1]*Naval Research Laboratory*, Washington, DC 20375
[2]*University of Crete*, Heraklion Crete, 71003, Greece
[3]*Institute of Electronic Structure and Laser (IESL), Foundation for Research and Technology Hellas (FORTH),* Heraklion Crete, 71110, Greece

**S1. Sample preparation: different samples used in this study.**

Three different samples were used in this study; an optical microscope image of each is presented in Figure S1. The majority of the data presented in this manuscript are from an exfoliated monolayer show in Figure S1a. To confirm the reproducibility of the effects reported in this manuscript, two other samples were also measured. These consisted of a monolayer that we exfoliated from a bulk crystal (Figure S1b) and a sample (Figure S1c) consisting of a large-area single-monolayer, grown at NRL by chemical vapor deposition (CVD). The CVD $WS_2$ was grown in a quartz tube furnace on $SiO_2$ (275nm). PTAS seeding particles are used to assist nucleation of the $WS_2$ on the substrate and are spun on the $SiO_2$ prior to growth. $WO_3$ (~1080mg) is placed at the center of the furnace with the $SiO_2$ directly above it, face-down. Sulfur is in a separate boat upstream in a cooler zone, slightly outside the central heating area. The growth is performed at atmospheric pressure with 100 sccm Ar and 10 sccm $H_2$ continuously flowing. The heater is ramped to 825˚C and held there for 10 minutes.

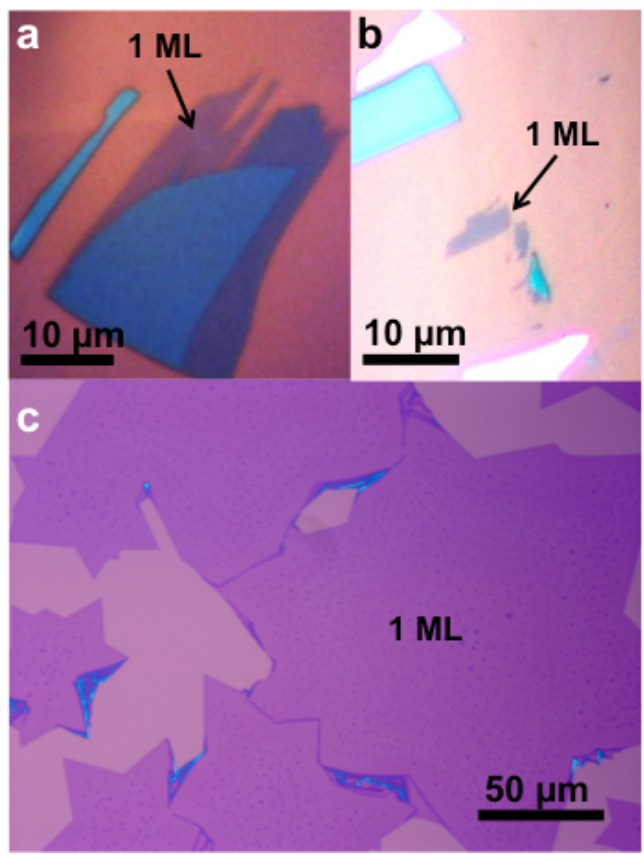

**Figure S1**. WS$_2$ samples used in this study. Different exfoliated flakes are shown in (a) and (b). (c) CVD grown WS$_2$.

## S2. Temperature dependent circular polarization.

Figure S2 is a compilation of raw polarization resolved PL spectra of the trion (X–) for all of the excitation energies and temperatures measured on the sample shown in Figure S1a. The incident laser light was polarized σ+, and the emission was analyzed for σ+ (red curves) and σ– (blue curves). The peak intensities have been normalized to the σ+ intensity at each temperature and are offset for clarity. The temperature resolved spectra at λ=594 nm and λ=532 nm shown here are also presented in the main text. When the sample is excited with σ– polarization, identical trends are observed. Figure 3c in the main text is derived from the spectra shown in Figure S2.

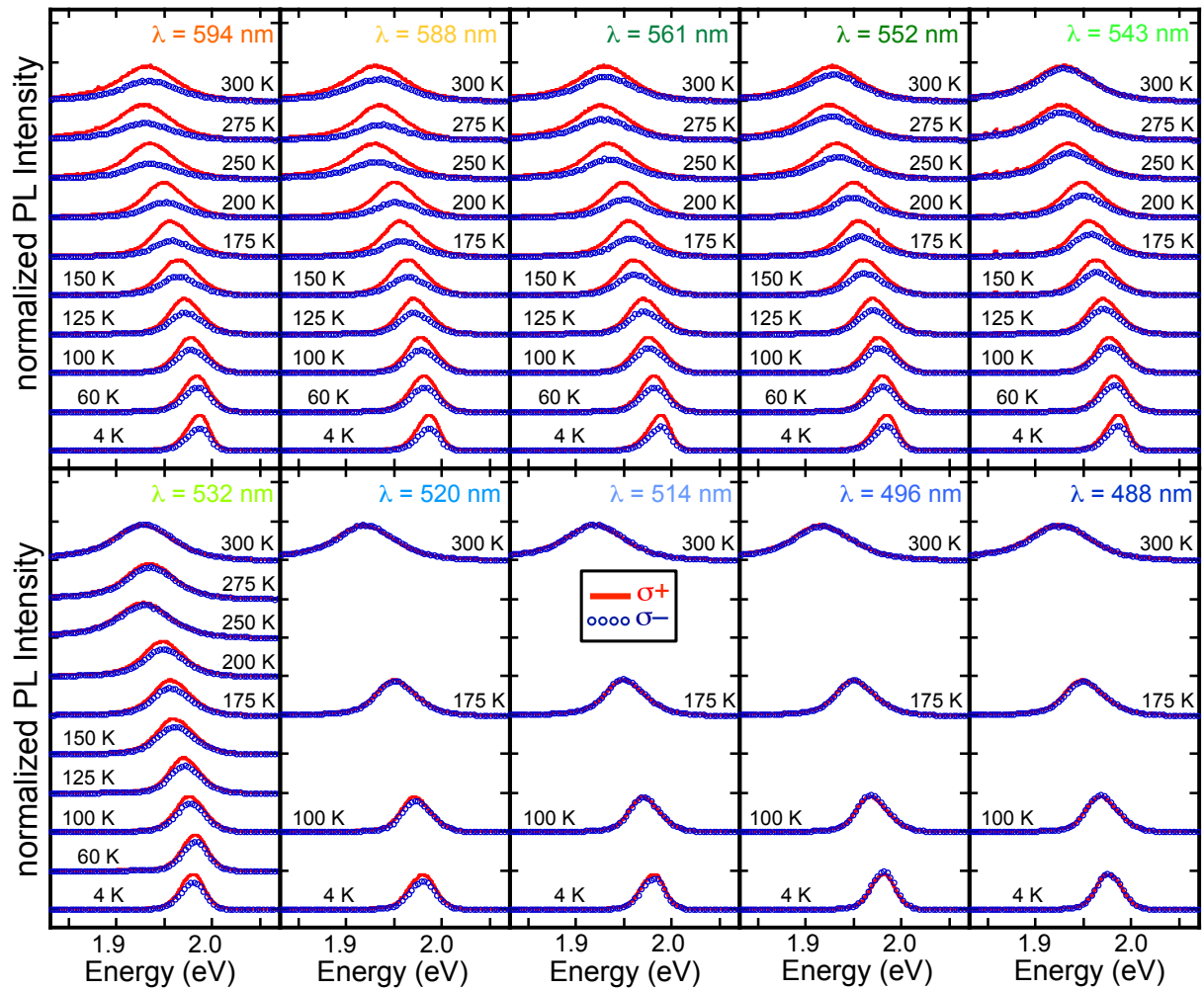

**Figure S2**. Photoluminescence analyzed for positive (σ+: solid, red trace) and negative (σ–: blue, open circles) helicity as a function of temperature of trion (X–) for 10 different excitation energies (indicated on each panel). At each temperature, the spectra are normalized to the σ+ intensity and offset for clarity.

## S3. Rate equation framework.

Within the rate equation framework, the time evolution of population of an exciton in a particular valley, K or K', can be described by the equations:

$$\frac{dN_K}{dt} = g_K - \frac{N_K}{\tau_r} - \frac{N_K - N_{K'}}{\tau_s}, \qquad (1)$$

$$\frac{dN_{K'}}{dt} = g_{K'} - \frac{N_{K'}}{\tau_r} + \frac{N_K - N_{K'}}{\tau_s}. \qquad (2)$$

Here, $g_{K/K'}$ denotes the optical pumping rate of the K or K' valley, $\tau_r$ is the radiative recombination time, and $\tau_s$ is the spin scattering time, or equivalently in this case the intervalley scattering time. In the main text, rates are used instead of times: the radiative recombination rate $\alpha = 1/\tau_r$, and the spin (or valley) scattering rate is $\beta = 1/\tau_s$. Under steady state conditions $dN_{K/K'}/dt = 0$. We define the polarization as $P = [N_K - N_{K'}]/[N_K + N_{K'}]$, which maps to the experimentally measured quantity $P = [I(\sigma+) - I(\sigma-)]/[I(\sigma+) + I(\sigma-)]$, where $I(\sigma\pm)$ is the emission intensity analyzed for positive (negative) helicity. The polarization can then solved in terms of the recombination and intervalley scattering time as

$$P = \frac{P_0}{\left(1 + 2 \cdot \frac{\tau_r}{\tau_s}\right)}. \qquad (3)$$

Here, $P_0 = [g_K - g_{K'}]/[g_K + g_{K'}]$, is the initial polarization of the system. In the main text instead of these characteristic times, we use the rates $\alpha$ and $\beta$. Equations (1) and (2) therefore become

$$\frac{dN_{K/K'}}{dt} = g_{K/K'} - \alpha N_{K/K'} \mp \beta(N_K - N_{K'}). \qquad (4)$$

As the temperature is increased, the recombination becomes a combination of the radiative and non-radiative recombination rates and $\alpha \rightarrow \alpha + A$, where $A$ is the non-radiative recombination rate. The polarization then becomes

$$P = \frac{P_0}{\left(1 + 2 \cdot \frac{\beta}{\alpha + A}\right)}. \qquad (5)$$

Figure S3 schematically shows the recombination and scattering processes considered.

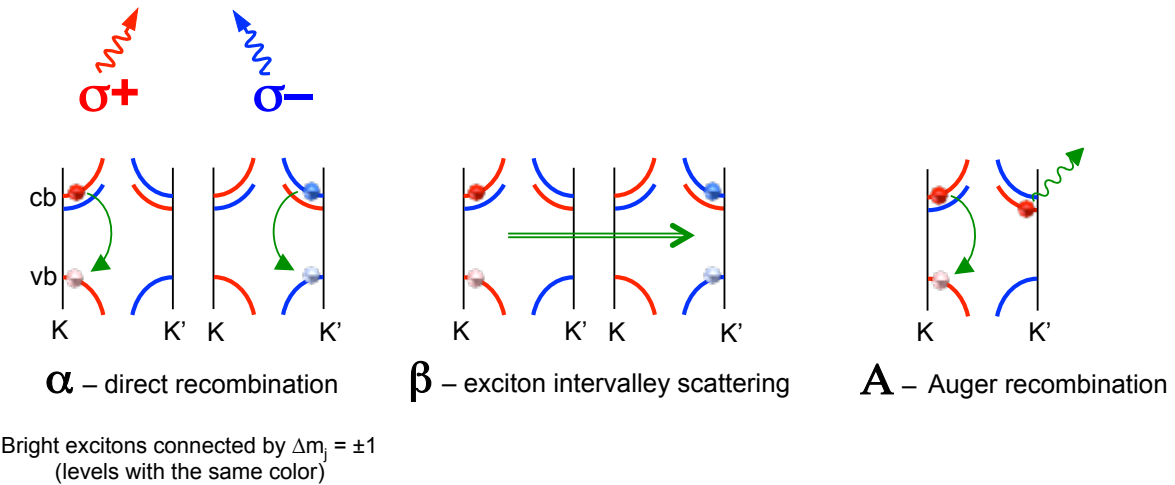

**Figure S3**. Schematic diagram of the recombination and scattering processes involved in our analysis. These include the direct recombination process of an electron-hole pair, the intervalley scattering of an exciton and the non-radiative, 3-particle Auger process.

**S4. Transition from Neutral Exciton to Trion.**

In figure 1 and figure 2 of the main text, spectra from the isolated neutral exciton and trion are presented. The system evolves smoothly from neutral exciton to trion, however, and these are only the endpoints. In the intermediate regime, peaks from both the neutral exciton and trion are visible in the PL and reflectivity spectra. Figure S4 shows the PL (figure S4a) and reflectivity (figure S4b) in this intermediate regime. At low temperature, with a low excitation power, the PL clearly shows both the neutral exciton and the trion. And as is the case for these peaks when they are isolated, the polarization of the neutral exciton is half that of the trion. The reflectivity at 4 K also shows both peaks and has a separation of 33 meV, the binding energy of the trion.

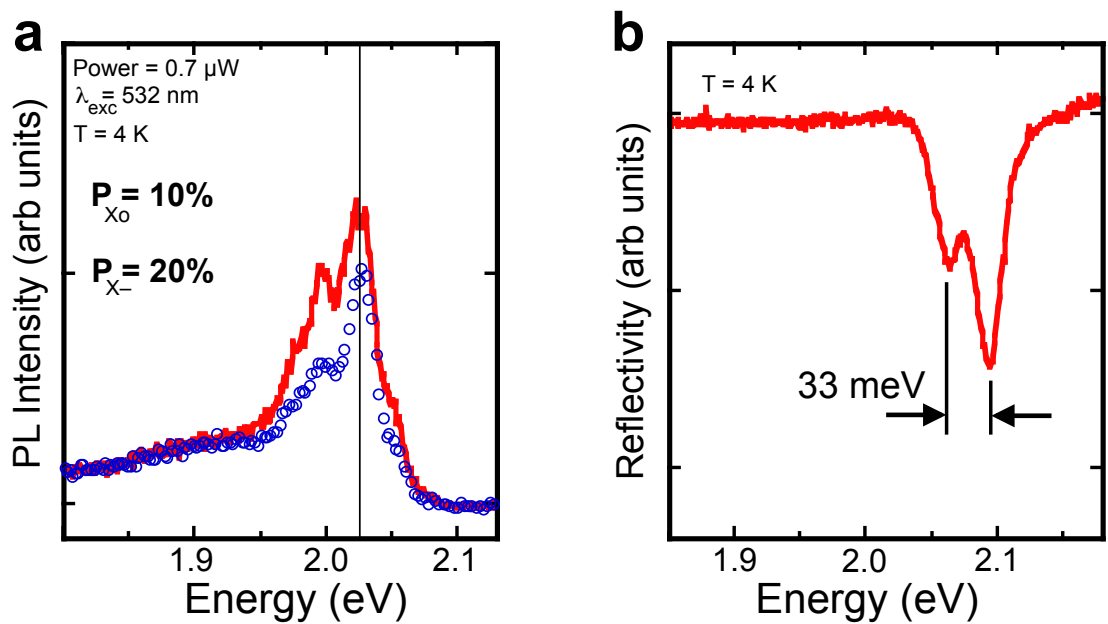

**Figure S4**. (a) Photoluminescence analyzed for positive ($\sigma+$: solid, red trace) and negative ($\sigma-$: blue, open circles) helicity at low temperature and power. The trion and neutral exciton are both clearly visible, and with a polarization ratio of 2, as is seen in the spectra where the trion and exciton are isolated. (b) Differential reflectivity at low temperature also showing the coexistent trion and neutral exciton states. The energy difference between these peaks is representative of the trion binding energy.